\documentclass[11pt]{article}
\usepackage[pdftex]{graphicx}
\pdfoutput=1

\newcommand{\We}{\mathrm{We}}
\renewcommand{\Re}{\mathrm{Re}}

\begin{document}
\title{Primary Atomization of a Liquid Jet in Crossflow}
\author{Sandeep Rana and Marcus Herrmann\\ \\\vspace{6pt} School for Engineering of Matter, Transport and Energy \\
Arizona State University, Tempe, AZ 85287, USA}
\maketitle
\begin{abstract} 
In this fluid dynamics video, we present a visualization of the primary atomization of a turbulent liquid jet injected into a turbulent gaseous crossflow. It is based on a detailed numerical simulation of the primary atomization region of the jet using a finite volume, balanced force, incompressible LES/DNS flow solver coupled to a Refined Level Set Grid (RLSG) solver to track the phase interface position. The visualization highlights the two distinct breakup modes of the jet: the column breakup mode of the main liquid column and the ligament breakup mode on the sides of the jet and highlights the complex evolution of the phase interface geometry.
\end{abstract}
\section{Introduction}
This {fluid dynamics video} presents a visualization of the primary atomization region of a turbulent liquid jet injected into a fast moving gaseous crossflow. This atomization scenario is commonly applied in gas turbine engines and augmentors where a liquid fuel jet is injected and broken up into a large number of small scale drops to form a spray.
Predicting the resulting spray drop size distribution is one of the outstanding challenges in atomizing multi-phase flows. Detailed simulations, as employed here, are starting to become a viable tool to study the physics of the atomization process and help develop models for predicting the resulting liquid spray.

The following sections summarize the methodology used in the flow simulation, give the computational geometry and operating conditions, and discuss briefly the flow features seen in the video. A detailed description of all these topics can be found in \cite{Herrmann08a,Herrmann10a,Herrmann10b,Herrmann10c,Herrmann10d}.

\section{Simulation Methodology}
The flow in the unsteady, incompressible, immiscible, two-fluid system is described by the Navier-Stokes equations augmented by a singular surface tension force term that is active only at the location of the phase interface. We describe the motion of the phase interface by solving a level set equation, such that the level set scalar is equal to the signed distance to the phase interface. Assuming that material properties like density and viscosity are constant in each phase, we follow a single fluid approach, by letting the density and viscosity jump at the phase interface, thereby assuming the phase interface to be a material discontinuity.

To keep track of the position and motion of the phase interface, we solve all level set related equations using the Refined Level Set Grid (RLSG) on a separate, equidistant Cartesian grid using a dual-narrow band methodology for efficiency. This so-called $G$-grid is overlaid onto the flow solver grid on which the Navier-Stokes equations are solved and can be independently refined, providing high resolution of the tracked phase interface geometry. Details of the resulting method and extensive verification results are reported in \cite{Herrmann08a}.

The flow solver used to solve the incompressible two-phase Navier-Stokes equations on unstructured grids using a finite volume balanced force algorithm is Cascade Technologies' CDP. In the single phase regions, the employed scheme conserves the kinetic energy discretely and turbulence is modeled using a dynamic Smagorinsky LES model.  However, none of the terms arising from filtering the phase interface, like subfilter surface tension or subfilter liquid volume fraction transport, are modeled. The approach instead relies on resolving all relevant scales at the phase interface, thus reverting to a DNS there \cite{Herrmann10a}. 
The flow solver CDP and the {RLSG} interface tracking software are coupled using the parallel multi-code coupling library CHIMPS \cite{Herrmann08a,Alonso06}.

Finally, resolving the entire phase interface geometry by tracking the phase interface associated with each atomized drop quickly becomes prohibitively expensive. Instead, we follow a multi-scale coupled Eulerian/ Lagrangian procedure in that we track the phase interface by the Eulerian level set method in the near injector primary atomization region and transfer broken-off, nearly spherical liquid structures into a Lagrangian point particle description \cite{Herrmann10b}. In the Lagrangian description, full two-way momentum coupling between the drop and continuous phase is used \cite{Moin06}, including a stochastic secondary atomization model \cite{Apte03}. 

\section{Computational Geometry and Operating Conditions}

The case analyzed  is one studied experimentally by Brown \& McDonell \cite{Brown06}. Table \ref{Tab:char} summarizes the operating conditions and resulting characteristic numbers. The analyzed case lies at the regime transition from column breakup to shear breakup \cite{Wu97} and thus is expected to exhibit different atomization mechanisms simultaneously.

\begin{table}[htdp]
\small
\begin{center}
\begin{tabular}{rc}
\hline
jet exit diameter $D$ [mm]& 1.3  \\
momentum flux ration $q$ & 6.6 \\
crossflow Weber number $\We_c$ & 330 \\
jet Weber number $\We_j$ & 2178  \\
crossflow Reynolds number $\Re_c$ & 5.7e5 \\
jet Reynolds number $\Re_j$ & 14079  \\
\hline
\end{tabular}
\caption{Operating conditions and characteristic numbers \cite{Herrmann10a}.\label{Tab:char}}
\end{center}
\end{table}%

\begin{figure}
 \begin{center}
\includegraphics[width=0.9\textwidth]{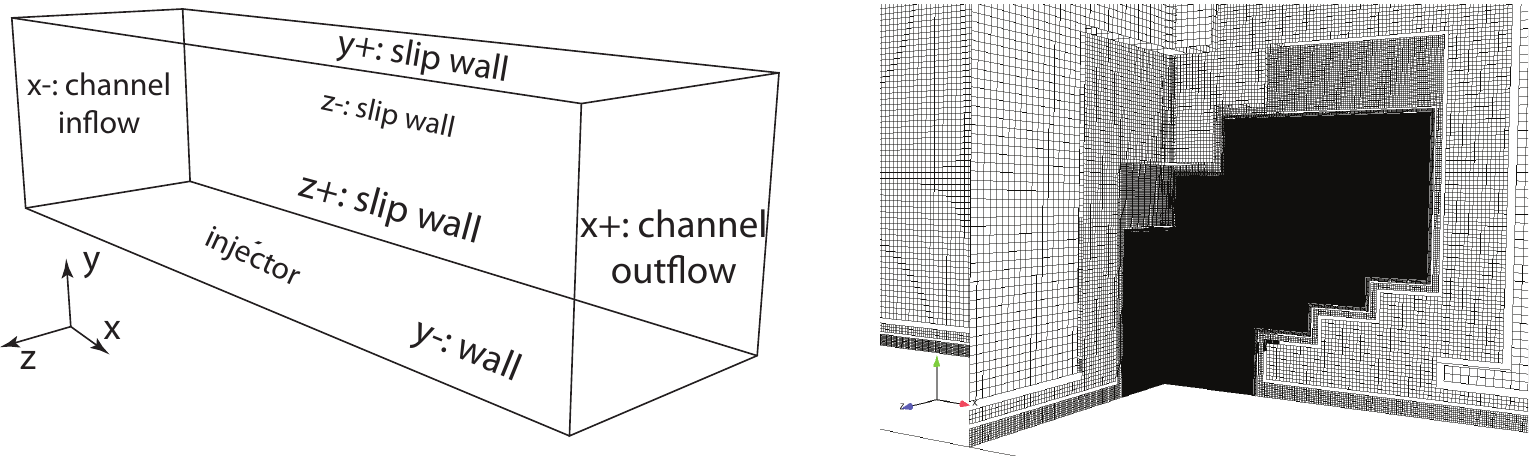}
\caption{\label{Fig:bc}Computational domain and boundary conditions (left) and mesh detail near the injector (right) \cite{Herrmann10a}.}
 \end{center}
\end{figure}
Figure \ref{Fig:bc} depicts the computational domain and the used boundary conditions as well as a zoom into the near-injector region to show the mesh detail used in the simulations. The computational domain is ($-25D \ldots 50D \times 0 \ldots 25D \times -10D \ldots 10D$) in dimension.

The injector geometry used in the experiments consists of a long initial pipe section of diameter 7.49mm, followed by an $138^o$ angled taper section, followed by a short pipe section of diameter $D$ with $L/D = 4$, whose exit is mounted flush with the lower channel wall. Since experimental results indicate that  the specifics of the liquid velocity distribution in the injector exit plane can have a large impact on the resulting atomization process of the liquid jet  \cite{Brown06,Brown07a}, the influence of the injector geometry on the turbulent exit plane velocity profiles are included in the simulation \cite{Herrmann10a}.

The simulation results shown in the video are obtained using a constant grid spacing in the primary atomization region of 32 grid points per injector diameter $D$ in the flow solver, and 64 grid points per $D$ for the $G$-grid. The resulting mesh size is 21 million control volumes for the flow solver and theoretically 840 million nodes for the RLSG solver, of which a maximum of 13 million are active at any given time. This resolution is the medium resolution case of a grid refinement study reported in \cite{Herrmann10a}.

\section{Visualization Methodology}
The flow simulation results visualized in the video consist of the liquid/gas phase interface and the generated individual spray drops. To visualize the phase interface, a marching cubes algorithm is employed to triangulate the $G=0$ iso-surface of the level set scalar. The spray drops are visualized as spheres having the drops' diameter, triangulated by a standard recursive triangulation algorithm. The level of recursive refinement for each sphere's surface is set proportional to the drop's diameter, which larger drops being refined more than small scale drops.
The resulting triangulated surfaces are ray traced using the open source software package Blender \cite{Blender}, assigning the liquid the reflective and refracted properties of water, colored by a small amount of blue dye to increase contrast.

\section{Results}
The video highlights the two major breakup modes by which the injected liquid breaks up. In the first, termed bag breakup mode, large scale instabilities grow on the windward side of the liquid column, forming bag like structures blown out by crossflow gas, causing rupture of the bags, leaving ligament structures behind near the top of the jet that then continue to breakup into different sized drops.

In the second breakup mode, termed ligament breakup mode, ligaments are generated near the injector exit at the side of the liquid jet. These get stretched out into long filaments that eventually rupture generating a broad range of drop sizes close to the bottom channel wall.

The video furthermore shows the complex paths individual drops can take in the wake region of the jet, with some drops clearly traveling upstream after generation, and the complex geometry of the liquid/gas phase interface induced by instability modes and the liquid jet's turbulence.

\section*{Acknowledgments}
This simulation work was supported by CASCADE Technologies Inc. under NavAir SBIR N07-046 and United Technologies Research Center. The visualizations were supported in part by Arizona State University's High Performance Computing Initiative.

\section*{References}

\end{document}